\documentclass[letterpaper]{article}

\usepackage[T1]{fontenc}

\usepackage{geometry}
\usepackage{setspace}

\usepackage{graphicx}
\usepackage{float}
\newfloat{scheme}{htbp}{los}
\floatname{scheme}{Scheme}
\floatname{chart}{Chart}
\newfloat{graph}{htbp}{loh}

\usepackage[version = 4]{mhchem} % \ce{}
\usepackage{amsmath}
\usepackage{amssymb}
\usepackage{xcolor}
\usepackage[colorlinks, citecolor=blue]{hyperref} % hyperref should generally be last

\usepackage{authblk}

\author[1]{Chun-Jie Zhang*}
\author[1]{Bing Zhang*}
\author[1]{Dongliang Mao*}
\author[1]{Yapeng Wu}
\author[1,2,3]{Xiao-Ping Li}
\author[1,2,3]{Lei Wang$^\dag$}

\affil[1]{Research Center for Quantum Physics and Technologies, School of Physical Science and Technology, Inner Mongolia University, Hohhot, 010021, China}
\affil[2]{Key Laboratory of Semiconductor Photovoltaic Technology and Energy Materials at Universities of Inner Mongolia Autonomous Region, Inner Mongolia University, Hohhot 010021, China}
\affil[3]{Inner Mongolia Key Laboratory of Microscale Physics and Atomic Manufacturing, Inner Mongolia University, Hohhot 010021, China}

\title{Charge-Density-Wave Phase Selection by Janus-Induced Intrinsic Strain in Monolayer \ce{NbSSiAs2}}

\date{
 *These authors contributed equally to this work.\\[0.5em]
 $^\dag$Email: lwang@imu.edu.cn
}

\begin{document}

\maketitle

\begin{abstract}
Controlling phase selection among competing charge-density-wave (CDW) instabilities remains challenging in two-dimensional materials. Here, first-principles calculations show that Janus-induced intrinsic tensile strain redirects the off-M soft-mode tendency of \ce{NbS2} to the M point in \ce{NbSSiAs2}, selecting a $2 \times 2$ CDW reconstruction. Electron–phonon coupling analysis identifies momentum-selective coupling between Nb-derived states and a longitudinal acoustic mode as the origin of the M-point instability. The reconstructed phase hosts two nearly degenerate Nb-trimerized configurations whose relative stability is tuned by biaxial strain. Both configurations retain phonon-mediated superconductivity on the 6--7 K scale, indicating the coexistence of CDW order and superconductivity. Compressive strain favors the $1+3$-hollow configuration and induces a band-inverted, $\mathbb{Z}_2$-nontrivial state while preserving superconductivity. Together, these results identify Janus-induced intrinsic strain as an internal structural route for CDW phase selection, whereas external strain provides access to a regime in which CDW order, topology, and superconductivity coexist.
\end{abstract}

%%%%%%%%%%%%%%%%%%%%%%%%%%%%%%%%%%%%%%%%%%%%%%%%%%%%%%%%%%%%%%%%%%%%%
%% Main Text
%%%%%%%%%%%%%%%%%%%%%%%%%%%%%%%%%%%%%%%%%%%%%%%%%%%%%%%%%%%%%%%%%%%%%
\section{Keywords} 
two-dimensional Janus quantum materials, charge-density-wave, intrinsic strain, topological phase, superconductivity

\section{Introduction} 
Charge density waves (CDWs) are collective ordered states driven by electron--lattice coupling in low-dimensional quantum materials. A CDW transition induces a periodic modulation of the electronic and lattice structures \cite{re1,re2,re3,re4,re5,re6}. This reconstruction can modify local bonding, orbital hybridization, and electronic states near the Fermi level, thereby affecting metallicity, superconductivity, and band topology \cite{re7,re8,re9,re10}. However, phonon softening or structural instability indicates only a tendency toward CDW formation, rather than uniquely determining the low-temperature reconstructed phase \cite{re11,re12,re13,re14,re15,re16,re17}. In two-dimensional transition-metal dichalcogenides (TMDs), multiple symmetry-allowed condensation channels or local minima in the free-energy landscape may coexist, giving rise to competing CDW configurations \cite{re18,re19,re20,re21,re22}. Clarifying the mechanism of CDW phase selection is therefore essential for understanding how distinct CDW orders shape the resulting electronic phases.

Strain provides a direct route to manipulating CDW order because it couples to the lattice distortion accompanying charge modulation. Externally applied or substrate-induced strain has been shown to renormalize soft phonons and modify the stability of competing CDW wave vectors and reconstructed phases \cite{re23,re24,re25,re26,re27,re28,re29}. In real materials, however, strain can also arise internally from lattice mismatch, inequivalent bonding, or local structural distortions, forming spatially inhomogeneous intrinsic strain fields \cite{re30,re31,re32,re33,re34,re35,re36}. Because these fields are embedded in the local bonding and electronic structure, they may reshape the local free-energy balance among competing CDW configurations. Identifying such an intrinsic structural control parameter would provide a new route for engineering CDW states.

Janus two-dimensional materials provide a unique platform for exploring intrinsic structural control of CDW states. Due to the asymmetric substitution between the top and bottom atomic layers, Janus structures break the out-of-plane mirror symmetry and create inequivalent chemical environments on the two sides of the monolayer \cite{re37,re38,re39,re40,re41,re42,re43}. Previous studies of Janus materials have mainly focused on symmetry-breaking-induced electronic phenomena, such as Rashba spin splitting and valley-dependent physics \cite{re44,re45,re46,re47,re48}. Beyond these electronic effects, the asymmetric chemical environments can also alter lattice energetics and structural instabilities, providing an additional structural degree of freedom for tuning CDW energy landscapes \cite{re49,re50,re51,re52}. For Janus systems hosting CDW instabilities, a pivotal question is how such intrinsic structural asymmetry influences the CDW energy landscape and selects among competing reconstructed phases.

Here, we demonstrate how intrinsic lattice asymmetry controls CDW phase selection in Janus \ce{NbSSiAs2}. We show that Janus-induced intrinsic strain acts as an effective structural degree of freedom, redirecting the parent soft-mode instability toward a commensurate $\mathrm{M}$-point ordering vector and reshaping the CDW energy landscape. The resulting $2 \times 2$ CDW phase hosts two nearly degenerate Nb-trimerized configurations whose relative ordering can be reversed by external strain. Furthermore, the strain-selected CDW states reconstruct the low-energy electronic structure, enabling a transition to a $\mathbb{Z}_2$-nontrivial phase while retaining electron--phonon-mediated superconductivity above 6 K. These findings identify intrinsic Janus strain as an effective route for engineering intertwined charge order, topology, and superconductivity in two-dimensional quantum materials.

\section{Method}
First-principles calculations were performed within density functional theory (DFT) using the Perdew--Burke--Ernzerhof functional \cite{re53,re54}. Phonon dispersions and constrained density functional perturbation theory (cDFPT) calculations were performed using the QUANTUM ESPRESSO package, combined with electron--phonon coupling (EPC) and phonon self-energy analyses using the EPW and ELPHMOD codes, to reveal the microscopic origin of CDW instabilities \cite{re55,re56,re57,re58}. Spin--orbit coupling (SOC) effects were considered for both the undistorted Janus structure and reconstructed CDW phases. The structural energetics and electronic structures of the reconstructed $2 \times 2$ CDW phases were investigated using the Vienna \textit{ab initio} Simulation Package with the projector augmented-wave method \cite{re59}. A vacuum spacing of 20 \AA{} and dipole corrections were applied for Janus monolayers (see Note S1 in the Supporting Information; see also refs. \cite{re60,re61} therein). Transition pathways between competing CDW configurations were determined using the nudged elastic band (NEB) method \cite{re62}, while the dependence of the CDW energy landscape on exchange--correlation functionals was examined. Finite-temperature stability was evaluated by \textit{ab initio} molecular dynamics (AIMD) simulations. Wannier-based tight-binding Hamiltonians were constructed, and edge spectral functions were calculated using WannierTools \cite{re63}.

\section{Intrinsic-Strain-Driven CDW Instability}

\begin{figure}[htbp]
  \centering
  \includegraphics[width=1.0\linewidth]{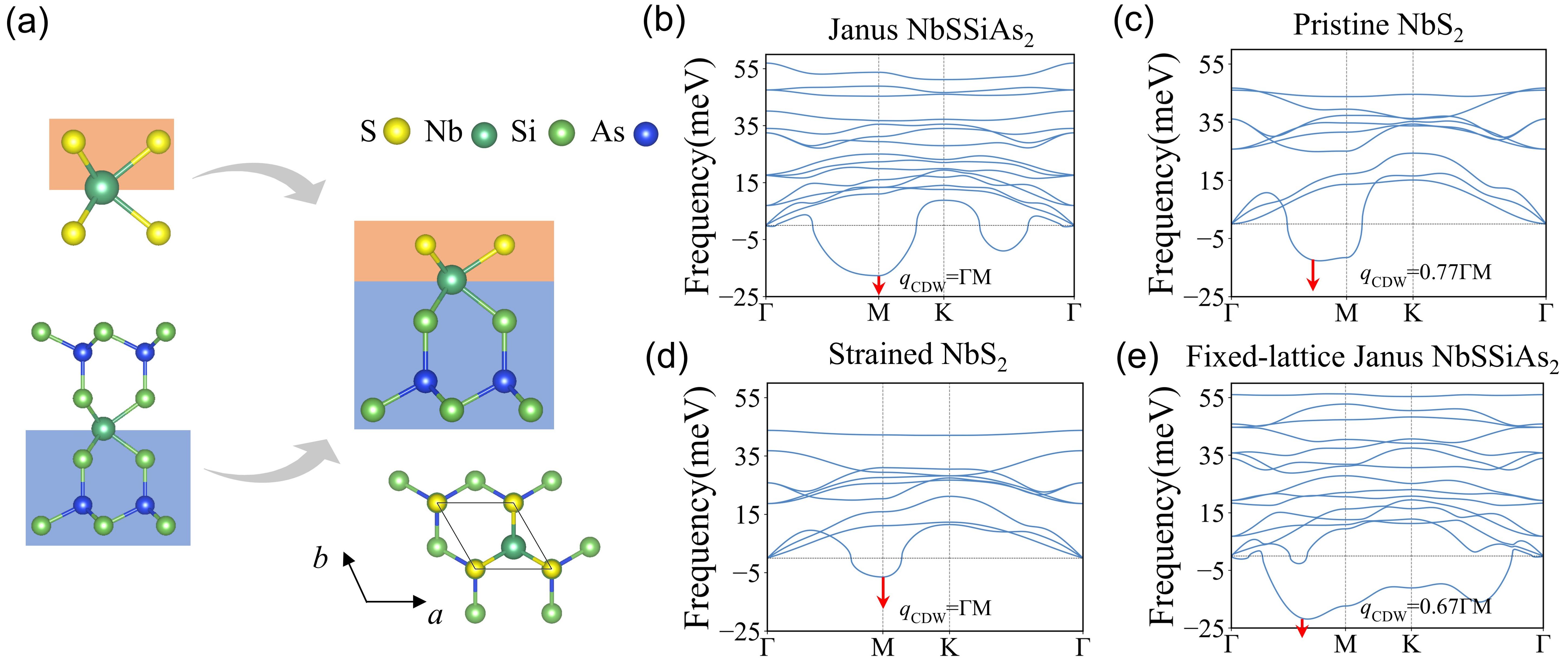}
 \caption{Janus asymmetry-induced intrinsic strain as a selector of commensurate CDW instability. (a) Schematic construction of Janus \ce{NbSSiAs2} with an asymmetric S--Nb--As--Si--As stacking sequence. (b--e) Phonon dispersions of (b) Janus \ce{NbSSiAs2}, (c) pristine \ce{NbS2}, (d) strained \ce{NbS2} with the in-plane lattice constant fixed to that of relaxed Janus \ce{NbSSiAs2}, and (e) fixed-lattice Janus \ce{NbSSiAs2} constrained to the pristine \ce{NbS2} lattice constant. Janus \ce{NbSSiAs2} exhibits a pronounced longitudinal acoustic (LA) phonon instability at $q_{\mathrm{CDW}} = \Gamma\mathrm{M}$, whereas pristine \ce{NbS2} shows an off-$\mathrm{M}$ instability at $q_{\mathrm{CDW}} \approx 0.77\Gamma\mathrm{M}$. Tensile strain alone shifts the instability toward $\mathrm{M}$, whereas the fixed-lattice Janus structure broadens the unstable region and moves the minimum to $q_{\mathrm{CDW}} \approx 0.67\Gamma\mathrm{M}$, indicating that intrinsic strain primarily controls the commensurate $2 \times 2$ CDW wave-vector selection.}
  \label{fig:fig1}
\end{figure}

A Janus \ce{NbSSiAs2} monolayer is derived from the CDW-active 1H-\ce{NbS2} parent structure by replacing one sulfur layer with an As--Si--As trilayer layer from the \ce{MA2Z4}-type \ce{NbSi2As4} structure (Figure~\ref{fig:fig1}a) \cite{re64,re65}. The resulting S--Nb--As--Si--As configuration breaks the out-of-plane mirror symmetry of the parent S--Nb--S trilayer, giving rise to intrinsic polarity and chemically distinct coordination environments above and below the Nb plane. The inequivalent coordination environments impose different in-plane lattice constraints on the Nb-centered lattice, driving it away from the equilibrium geometry of pristine \ce{NbS2}. Consequently, the in-plane lattice constant expands from 3.346 \AA{} in \ce{NbS2} to 3.592 \AA{} in \ce{NbSSiAs2}, corresponding to a 7.35\% intrinsic tensile pre-strain. Phonon calculations reveal a pronounced imaginary frequency in the longitudinal acoustic (LA) branch at $\mathrm{M}$ ($q_{\mathrm{CDW}} = \Gamma\mathrm{M}$), indicating a dominant commensurate $2 \times 2$ CDW instability that persists with SOC included (Figure~\ref{fig:fig1}b and Figure~S2).

To determine whether this $\mathrm{M}$-point instability is new or a shifted version of the parent \ce{NbS2} soft mode, we compared Janus \ce{NbSSiAs2} with pristine \ce{NbS2} using the same computational protocol. Pristine \ce{NbS2} exhibits a soft acoustic mode along $\Gamma\mathrm{M}$, with the lowest-frequency instability located at $q_{\mathrm{CDW}} \approx 0.77\Gamma\mathrm{M}$ (Figure~\ref{fig:fig1}c). Although the precise position of this soft mode can vary with computational details \cite{re66,re67}, this off-$\mathrm{M}$ instability points to an incommensurate distortion close to the previously reported near-$3 \times 3$ CDW tendency \cite{re8,re68}. By contrast, Janus \ce{NbSSiAs2} shows its soft mode at the $\mathrm{M}$ point, suggesting that the Janus reconstruction redirects the parent soft-mode tendency toward a commensurate $2 \times 2$ instability.

This wave-vector shift motivates a separate assessment of how Janus-induced out-of-plane polarity and intrinsic strain contribute to the CDW wave-vector selection. We therefore consider two complementary constrained reference structures to evaluate their relative roles. In the strain-only reference model, the \ce{NbS2} composition and symmetric S--Nb--S trilayer geometry are retained, but the in-plane lattice constant is fixed to that of relaxed Janus \ce{NbSSiAs2}. In the fixed-lattice polar Janus model, the lower S layer of \ce{NbS2} is replaced by the As--Si--As motif, whereas the in-plane lattice constant is kept at that of pristine \ce{NbS2}. As shown in Figures~\ref{fig:fig1}d and~\ref{fig:fig1}e, the strain-only \ce{NbS2} model shifts the lowest phonon instability from the parent off-$\mathrm{M}$ wave vector to the $\mathrm{M}$ point, reproducing the wave-vector selection of Janus \ce{NbSSiAs2}. By contrast, the fixed-lattice Janus model shifts the instability toward $q_{\mathrm{CDW}} \approx 0.67\Gamma\mathrm{M}$ and produces a broader imaginary branch.

\begin{figure}[htbp]
  \centering
  \includegraphics[width=1.0\linewidth]{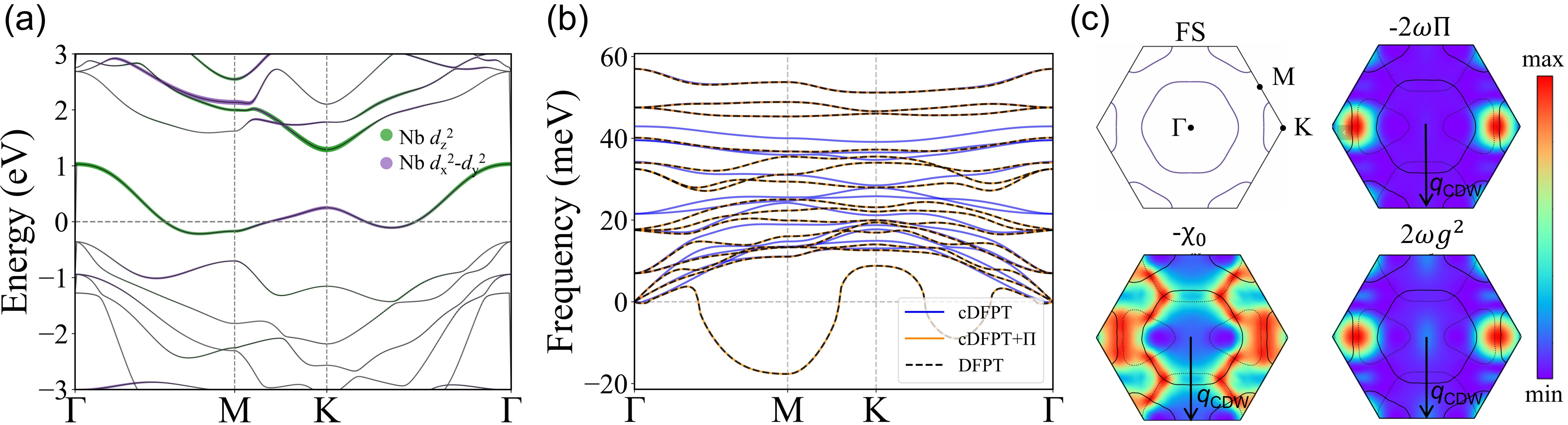}
  \caption{Electronic origin of the $\mathrm{M}$-point phonon instability. (a) Nb-$d$ fat-band electronic structure near the Fermi level. (b) Phonon dispersions from DFPT, cDFPT, and self-energy-corrected cDFPT. (c) Fermi surface together with momentum-resolved fluctuation diagnostics of the LA-mode softening at  $q_{\mathrm{CDW}} = \Gamma\mathrm{M}$. The $k$-resolved distributions of the phonon self-energy $2\omega\Pi$, bare electronic susceptibility $\chi_0$, and electron--phonon coupling strength $2\omega g^2$ are shown on a color scale. Solid and dashed contours indicate the original and $q_{\mathrm{CDW}}$-translated Fermi surfaces, respectively.}
  \label{fig:fig2}
  \end{figure}

To elucidate the electronic origin of the $\mathrm{M}$-point phonon instability, we performed constrained density functional perturbation theory (cDFPT) \cite{re56}, taking the Nb-derived metallic bands near the Fermi level as the active subspace (Figure~\ref{fig:fig2}a). The electronic screening arising from transitions within the Nb $d$ manifold was excluded, while all remaining screening channels were retained. As shown in Figure~\ref{fig:fig2}b, excluding the screening contribution from the Nb $d$ manifold completely removes the imaginary LA phonon mode at $\mathrm{M}$, indicating that the softening is primarily associated with the low-energy electronic response of the Nb $d$ states.

The electronic contribution to the phonon softening is quantified through the phonon self-energy $\Pi$. The renormalized dynamical matrix can be expressed as \cite{re69}:
\begin{equation}
\omega^2_{q\mu\nu} = \omega^2_{q\nu}\delta_{\mu\nu} + 2\omega_{q\mu}\omega_{q\nu}\Pi_{q\mu\nu}
\end{equation}
where $\mu$ and $\nu$ denote phonon branches and $\delta_{\mu\nu}$ is the Kronecker delta. $\omega_{q\nu}$ denotes the bare phonon frequency obtained from cDFPT, whereas $\Pi_{q\mu\nu}$ represents the electronic self-energy correction arising from electron–phonon coupling. $\Pi_{q\mu\nu}$ quantifies how specific electronic states renormalize the phonon spectrum. 

To identify the microscopic electronic channel responsible for the $\mathrm{M}$-point phonon softening, we performed momentum-resolved fluctuation diagnostics for $\Pi_{q\mu\nu}$ evaluated at  $q_{\mathrm{CDW}} = \Gamma\mathrm{M}$. $\Pi_{q\mu\nu}$ can be decomposed as:
\begin{equation}
\Pi_{q\mu\nu} = \sum_{kmn} g^{\mu *}_{kn,k+qm} g^{\nu}_{kn,k+qm} \chi^{0}_{qkmn}
\end{equation}
where $g^{\nu}_{kn,k+qm}$ denotes the electron-phonon coupling matrix element connecting electronic states $|kn\rangle$ and $|k+qm\rangle$ and $\chi^{0}_{qkmn}$ is the corresponding bare electronic susceptibility. As shown in Figure~\ref{fig:fig2}c, the k-resolved contributions of the $2\omega\Pi$, $g^2$, and $\chi_0$ at the fixed CDW wave vector $q_{\mathrm{CDW}}$. The $2\omega\Pi$ contribution exhibits pronounced hot spots associated with K/K$^\prime$ intervalley scattering, which closely correlate with the distribution of $2\omega g^2$, whereas $\chi_0$ shows much weaker enhancement and lacks comparable momentum selectivity \cite{re56,re64}. Thus, the $\mathrm{M}$-point phonon instability cannot be explained by conventional Fermi-surface nesting alone, but is primarily driven by strongly momentum-selective coupling between the Nb $d$ bands and the soft LA phonon mode. This picture is consistent with the EPC-driven CDW mechanism reported in monolayer $\alpha_2$-\ce{NbSi2As4} \cite{re65}, while the broader distribution of $\chi_0$ in the present system further underscores the dominant role of the momentum-dependent electron--phonon vertex.

\section{Strain-Tunable Bistable CDW States}

\begin{figure}[htbp]
  \centering
  \includegraphics[width=0.85\linewidth]{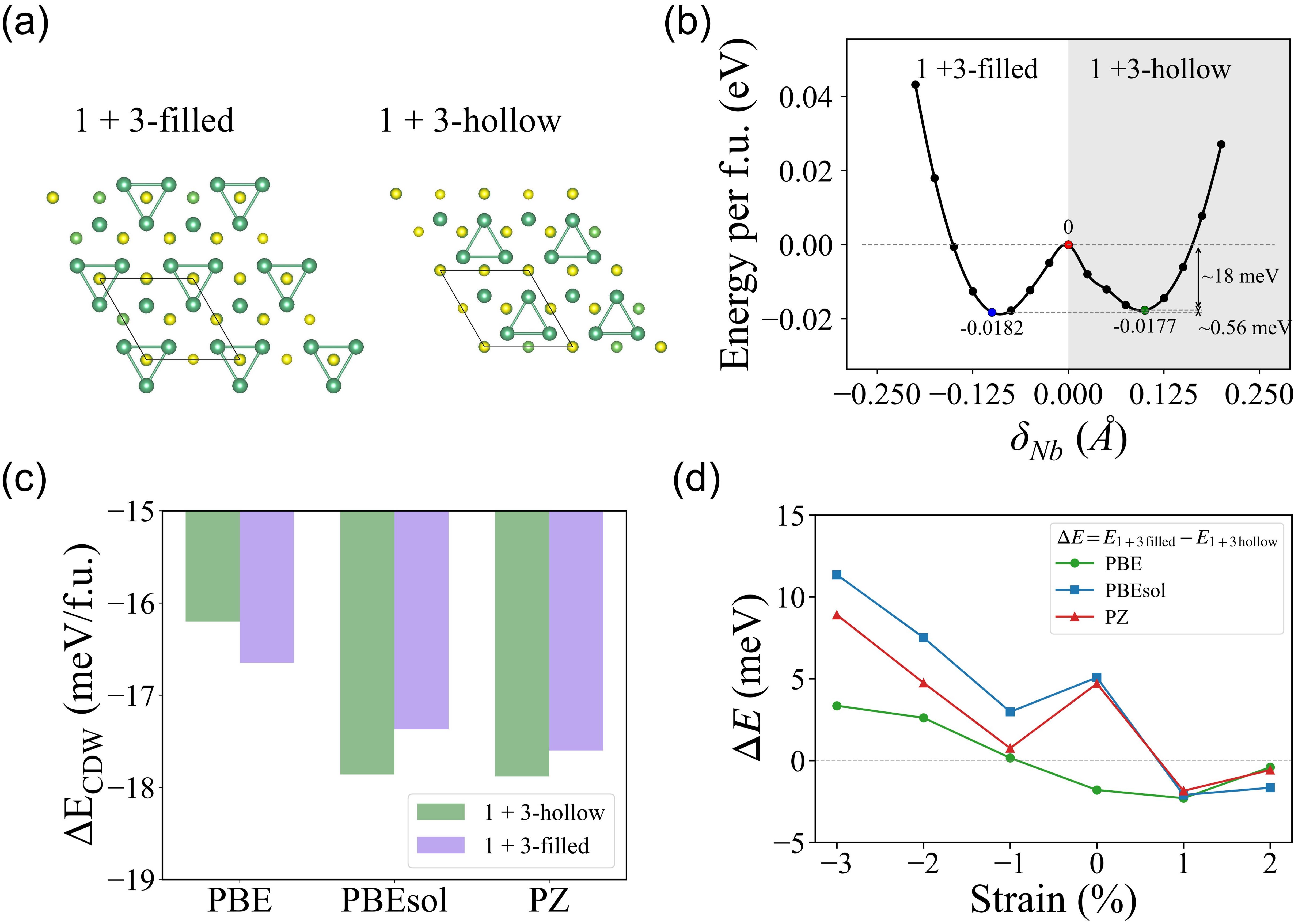}
  \caption{Bistable CDW configurations and strain-tunable phase competition in Janus \ce{NbSSiAs2}. (a) Relaxed $2\times2$ CDW structures of $1+3$-filled and $1+3$-hollow with a $1+3$ Nb-trimerized motif. (b) NEB energy profile connecting the two CDW states along the Nb-trimer distortion coordinate $\delta_{\mathrm{Nb}}$, revealing a double-well energy landscape with nearly degenerate minima. (c) CDW condensation energies of $1+3$-filled and $1+3$-hollow calculated using different exchange--correlation functionals. (d) Strain dependence of $\Delta E = E_{\text{1+3-filled}} - E_{\text{1+3-hollow}}$, showing strain-induced reversal of the relative stability between the competing CDW states.}
  \label{fig:fig3}
\end{figure}

The $\mathrm{M}$-point phonon instability supports two distinct condensation pathways. Condensation of a single $\mathrm{M}$-point mode produces a stripe-like $1\times2$ modulation, whereas simultaneous condensation of symmetry-equivalent $\mathrm{M}$-point modes gives rise to a $2\times2$ Nb-trimerized phase (Figure~S3). Total-energy relaxations show that the $2\times2$ reconstruction is lower in energy than the competing $1\times2$ structure by 31\,meV/f.u., identifying it as the favored CDW ground state associated with the $\mathrm{M}$-point instability. Structural relaxations further reveal two locally stable Nb-trimerized configurations within this $2\times2$ phase. Both exhibit a common $1+3$ clustering motif consisting of a compact Nb trimer and an isolated Nb atom within the $2\times2$ supercell. Depending on whether the center of the trimer is located at a filled or hollow site of the parent Nb lattice, the two configurations are denoted as $1+3$-filled and $1+3$-hollow, respectively (Figure~\ref{fig:fig3}a).

To characterize the energy landscape connecting these competing configurations, we calculated the minimum-energy path between the fully relaxed $1+3$-filled and $1+3$-hollow structures using the NEB method. The energy profile along the NEB path was projected onto the Nb-trimer distortion coordinate $\delta_{\mathrm{Nb}}$, defined as the collective radial displacement of the three Nb atoms within each $C_3$-symmetric trimer, with negative and positive values corresponding to $1+3$-filled and $1+3$-hollow, respectively (Figure~\ref{fig:fig3}b). The minimum-energy path passes through a high-symmetry saddle point near $\delta_{\mathrm{Nb}} = 0$ and connects two minima corresponding to $1+3$-filled and $1+3$-hollow, forming a double-well energy landscape. The two minima differ by only 0.56\,meV/f.u., while the saddle point is located approximately 18\,meV/f.u. above the lower-energy minimum. The coexistence of two nearly degenerate minima separated by a finite energy barrier establishes a bistable CDW landscape.

To assess the robustness of this bistable energy landscape against the choice of exchange-correlation functional, we recalculated the CDW condensation energies using several functionals. Both $1+3$-filled and $1+3$-hollow remain energetically favored over the undistorted reference phase for all functionals. Although their relative energetic ordering varies with the functional, the energy separation between the two phases remains small, supporting the persistence of a nearly degenerate bistable CDW landscape (Figure~\ref{fig:fig3}c). Phonon calculations reveal no imaginary frequencies for either configuration 
(Figure~S4), confirming their dynamical stability. AIMD at 300~K further show no tendency toward relaxation into the undistorted phase within the accessible simulation timescale (Figure~S5).

\begin{figure}[H]
  \centering
  \includegraphics[width=0.8\linewidth]{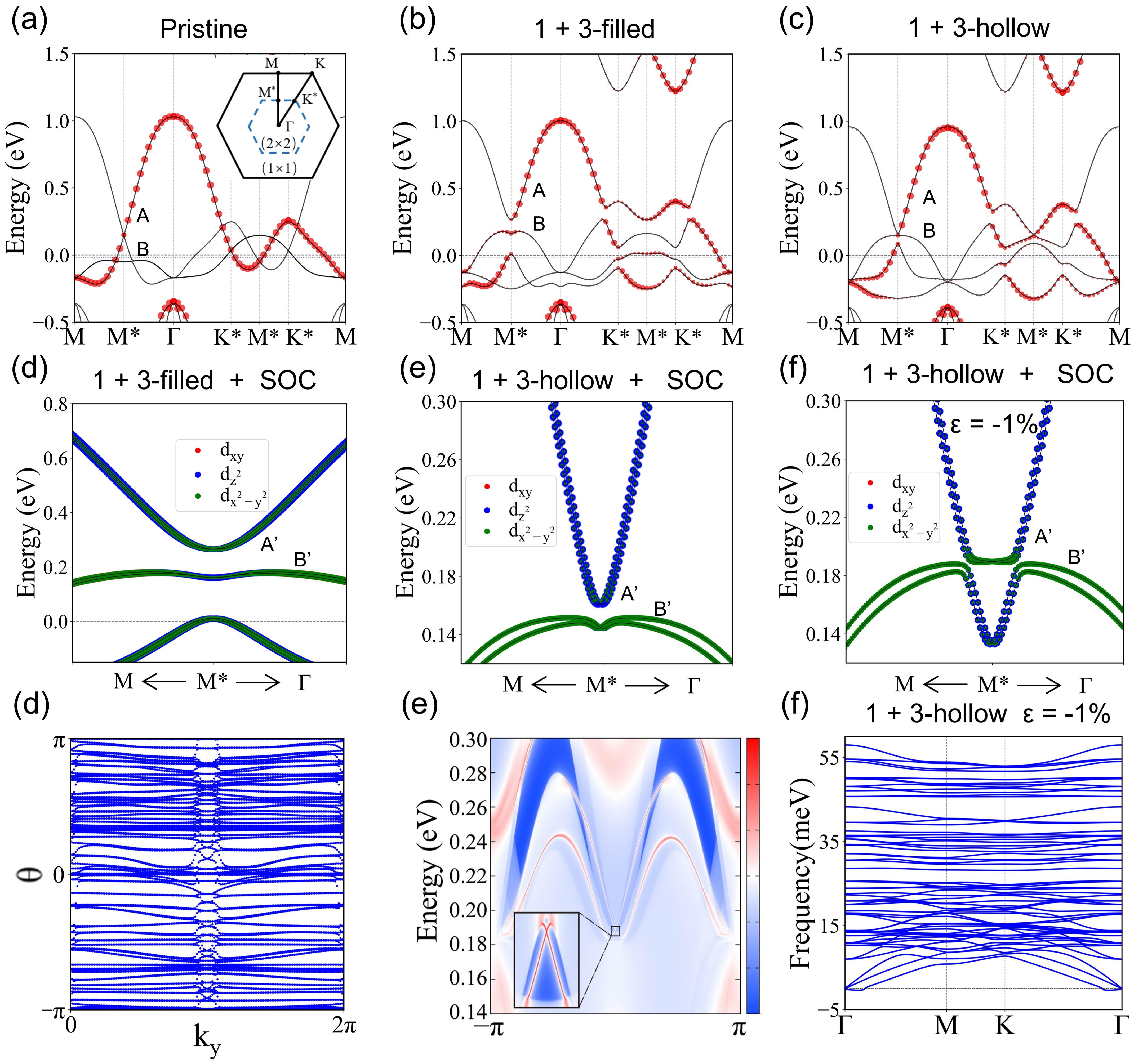}
  \caption{Strain-induced topological phase transition in the CDW state. (a--c) Unfolded band structures of pristine, $1+3$-filled, and $1+3$-hollow configurations, respectively. The black curves represent the folded bands of the $2\times2$ CDW supercell, while the red markers denote the unfolded spectral weights projected onto the primitive Brillouin zone, with marker size proportional to the spectral intensity. (d--f) SOC-included orbital-projected band structures of $1+3$-filled, unstrained $1+3$-hollow, and $1\%$ compressively strained $1+3$-hollow configurations, respectively, showing the evolution of the Nb $d_{z^2}$- and $d_{x^2-y^2}$-derived states near $\mathrm{M}^*$. (g) Wilson-loop spectrum of the strained $1+3$-hollow configuration, revealing a nontrivial $\mathbb{Z}_2$ invariant. (h) (010)-surface spectral function calculated from the Wannier-interpolated Hamiltonian, gapless edge states traversing the bulk gap. (i) Phonon spectrum of the strained $1+3$-hollow configuration, confirming its dynamical stability.}
  \label{fig:fig4}
  \end{figure} 

Given the near degeneracy of the two $2\times2$ CDW configurations, their energetic competition is expected to be sensitive to external perturbations. We therefore investigate whether externally applied biaxial strain can modulate, and potentially reverse, the energetic preference between $1+3$-filled and $1+3$-hollow. As shown in Figure~\ref{fig:fig3}d, the relative energy difference $\Delta E = E_{\text{1+3-filled}} - E_{\text{1+3-hollow}}$ was evaluated under biaxial strains ranging from $-3\%$ to $+2\%$, with negative strain denoting compression. PBE, PBEsol, and PZ consistently predict a pronounced strain dependence of $\Delta E$ and a reversal of the relative phase stability. Despite quantitative differences in their zero-strain energetic ordering and predicted crossover strain, all three functionals demonstrate that the competition between the two nearly degenerate CDW states can be effectively controlled by biaxial strain.

\section{Strain-Induced Topological CDW Phase}

The two competing CDW phases exhibit distinct low-energy electronic structures despite their nearly degenerate total energies. To elucidate the electronic consequences associated with different CDW configurations, we analyze their unfolded band structures within the PBE framework. The $2\times2$ CDW order folds the Brillouin zone and substantially modifies the electronic states near the folded $\mathrm{M}^*$ point (Figures~\ref{fig:fig4}a--c, Figure~S6). In particular, two CDW-derived bands (A and B) dominate the near-gap electronic structure and are closely associated with the orbital evolution underlying the topological transition.

At zero strain, both CDW configurations are topologically trivial, as confirmed by Wilson-loop calculations with $\mathbb{Z}_2 = 0$ (Figure~S7). Despite their identical topological character, their reconstructed electronic structures near $\mathrm{M}^*$ exhibit distinct band evolution. For the $1+3$-filled configuration, the Nb $d_{z^2}$ and $d_{x^2-y^2}$-derived states maintain a conventional ordering with a relatively larger energy separation (Figure~\ref{fig:fig4}d). In contrast, the $1+3$-hollow configuration shows a reduced separation between bands $A'$ and $B'$ near $\mathrm{M}^*$, suggesting that this configuration is closer to the band-inversion regime (Figure~\ref{fig:fig4}e). Upon applying $1\%$ compressive biaxial strain, the orbital characters of these two states exchange near $\mathrm{M}^*$, indicating a strain-induced band inversion (Figure~\ref{fig:fig4}f). Meanwhile, SOC opens a gap at the inverted bands, resulting in a gapped band-inverted electronic structure.

To further confirm the nontrivial topology, we calculate the Wilson-loop spectrum of $1+3$-hollow under $1\%$ compressive strain. Unlike the trivial state, the Wilson-loop spectrum exhibits an odd number of crossings of the Wannier charge centers with an arbitrary reference line, yielding $\mathbb{Z}_2 = 1$ (Figure~\ref{fig:fig4}g). Consistent with the bulk--boundary correspondence, the edge spectral function reveals topological edge states traversing the bulk gap (Figure~\ref{fig:fig4}h). Moreover, the absence of imaginary phonon frequencies confirms the dynamical stability of the strain-induced phase (Figure~\ref{fig:fig4}i). Therefore, the $1+3$-hollow configuration under $1\%$ compressive strain realizes a dynamically stable topological CDW state. The distinct electronic structures of the competing CDW configurations highlight the role of CDW reconstruction in tuning the proximity to the $\mathbb{Z}_2$ topological transition.

\section{Robust Superconductivity Across the Strain-Tunable CDW Landscape}

\begin{figure}[htbp]
  \centering 
  \includegraphics[width=1.0\linewidth]{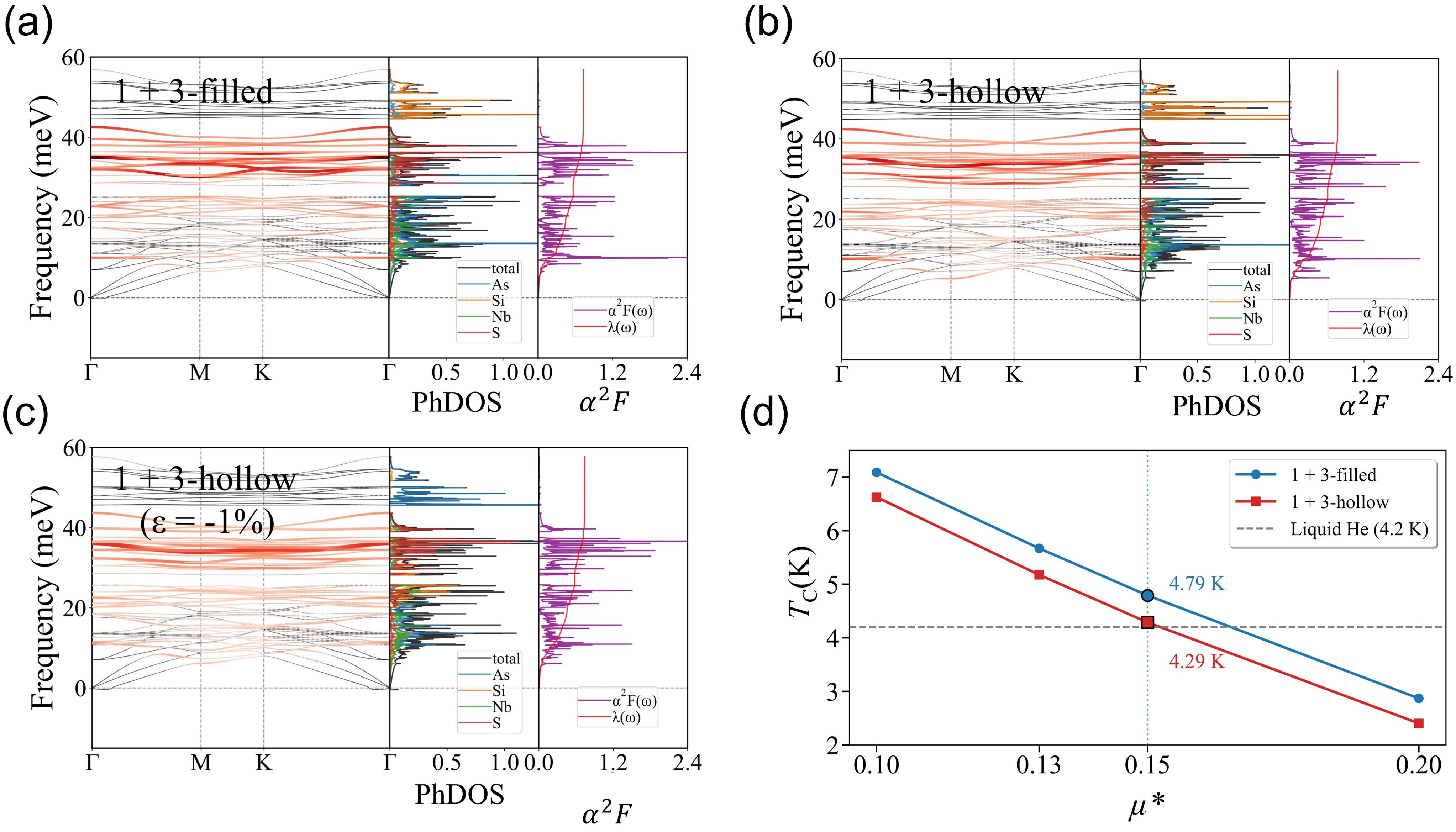}
  \caption{Phonon properties and superconducting transition temperatures of different phases. (a--c) Phonon properties of (a) $1+3$-filled, (b) $1+3$-hollow, and (c) compressively strained $1+3$-hollow phases. The phonon dispersions are shown together with the $\lambda_{q\nu}$-scaled phonon dispersions (red), where the intensity represents the mode-resolved electron--phonon coupling strength. The projected phonon density of states (PhDOS) illustrates the atomic contributions to the phonon modes, while the Eliashberg spectral function $\alpha^2F(\omega)$ and cumulative $\lambda(\omega)$. (d) Variation of the superconducting transition temperature $T_C$ with Coulomb pseudopotential $\mu^*$.}
  \label{fig:fig5}
  \end{figure}

Since CDW order often competes with superconductivity by reconstructing the Fermi surface and depleting low-energy electronic states, it is important to examine whether electron--phonon-mediated superconductivity survives in the competing CDW phases and in the strain-induced topological regime. The unfolded band structures in Figures~\ref{fig:fig4}b and~\ref{fig:fig4}c show that both reconstructed CDW phases retain finite spectral weight near the Fermi level, suggesting that the CDW reconstruction does not fully gap out the low-energy electronic states required for pairing. As shown in Figures~\ref{fig:fig5}a and~\ref{fig:fig5}b, the projected phonon density of states reveals that low-frequency vibrations are mainly associated with heavier Nb and As atoms, whereas higher-frequency modes originate from lighter S and Si atoms. These phonon features are reflected in the Eliashberg spectral function $\alpha^2F(\omega)$, with the dominant spectral weight located below 40\,meV, leading to the rapid accumulation of $\lambda(\omega)$ in this energy range. These results indicate that the CDW-reconstructed lattices retain effective electron--phonon coupling channels for superconducting pairing.

Using the Allen--Dynes formula with a Coulomb pseudopotential $\mu^* = 0.10$, the estimated superconducting transition temperatures are 6.63\,K and 7.09\,K for $1+3$-filled and $1+3$-hollow, respectively (Table~\ref{tbl:properties}). These values are comparable to those of representative Nb-based dichalcogenide superconductors, including 2H-\ce{NbS2} ($\sim$6\,K) and 2H-\ce{NbSe2} ($\sim$7.2\,K), showing that both competing CDW phases retain superconductivity on the characteristic temperature scale of established layered superconductors despite their distinct lattice registries and reconstructed electronic structures \cite{re67,re68}.

Importantly, a comparable superconducting scale is also preserved in the topological regime. For $1+3$-hollow under $1\%$ compressive biaxial strain, where the $\mathbb{Z}_2$-nontrivial band topology is realized, the calculated $T_C$ remains 6.21\,K (Figure~\ref{fig:fig5}c). Although slightly reduced compared with the unstrained $1+3$-hollow, this finite $T_C$ demonstrates that the strain-induced topological transition is not accompanied by a detrimental suppression of superconductivity. We further examine the sensitivity of $T_C$ to $\mu^*$. As shown in Figure~\ref{fig:fig5}d, even for $\mu^* = 0.15$, the estimated $T_C$ values remain above the liquid-helium temperature of 4.2\,K. These results indicate that Janus \ce{NbSSiAs2} is not simply a superconducting CDW material, but a strain-tunable coexistence platform in which CDW bistability, $\mathbb{Z}_2$ topology, and electron--phonon-mediated superconductivity are retained within the same reconstructed lattice landscape.

  \begin{table}[htbp]
  \caption{Structural and superconducting properties of the competing $1 + 3$-filled and strain-modulated $1 + 3$-hollow CDW phases.}
  \label{tbl:properties}
  \centering
  \begin{tabular}{llllll}
    \hline
    Phase & $a$ (\AA) & $T_C$ (K) & $\lambda$ & $\omega_{\text{log}}$ (K) & Reference \\
    \hline
    $1+3$-filled & 7.18 & 6.63 & 0.73 & 172.03 & This work \\
    $1+3$-filled & 7.18 & 7.09 & 0.78 & 160.45 & This work \\
    $1+3$-hollow ($\varepsilon = -1\%$) & 7.10 & 6.21 & 0.74 & 168.46 & This work \\
    2H-\ce{NbSe2} (Bulk) & - & 7.2 & - & - & Exp. \cite{re70} \\
    2H-\ce{NbS2} (Bulk) & - & 5.7 & - & - & Exp. \cite{re71} \\
    \hline
  \end{tabular}
\end{table}

\section{Conclusion}

In summary, we identify intrinsic Janus strain in \ce{NbSSiAs2} monolayer as a structural degree of freedom governing CDW phase selection. Unlike conventional strain tuning based on external mechanical perturbations, Janus structural asymmetry provides a built-in lattice degree of freedom that reshapes the CDW energy landscape. First-principles calculations demonstrate that this intrinsic tensile strain redirects the parent phonon instability toward a commensurate $\mathrm{M}$-point ordering vector, resulting in a $2 \times 2$ CDW reconstruction. The reconstructed phase landscape contains competing Nb-trimerized configurations with nearly degenerate energies. Moreover, CDW reconstruction modifies the electronic structure, inducing band inversion and a transition from a topologically trivial to a $\mathbb{Z}_2$-nontrivial phase while preserving robust electron–phonon-mediated superconductivity. These results establish Janus structural asymmetry as a promising strategy for regulating intertwined lattice, CDW, topological, and superconducting orders in two-dimensional quantum materials.

\section*{Acknowledgements}

This work was financially supported by the National Natural Science Foundation of China (Grants No. 12304165, No. 12304086 and No. 12564017), the Natural Science Foundation of Inner Mongolia Autonomous Region (Grant No. 2026QB016), the “Grassland Talents” project of the Inner Mongolia Autonomous Region (Grant No. 21200-242920), and the Startup Project of Inner Mongolia University (Grant No. 21200-5223733).

\section*{Supporting Information}

The following files are available free of charge.
\begin{itemize}
  \item SI.pdf: detailed computational methods, structural and electronic structure calculations, phonon stability analyses, \textit{ab initio} molecular dynamics simulations, band unfolding, and Wilson-loop-based topological characterizations.
\end{itemize}

%%%%%%%%%%%%%%%%%%%%%%%%%%%%%%%%%%%%%%%%%%%%%%%%%%%%%%%%%%%%%%%%%%%%%
%% Bibliography
%%%%%%%%%%%%%%%%%%%%%%%%%%%%%%%%%%%%%%%%%%%%%%%%%%%%%%%%%%%%%%%%%%%%%

\end{document}